\documentstyle[12pt]{ouriop}          
\begin{document}
\title{The distorted-wave impulse approximation for electron capture into
excited states}[Electron capture at intermediate energies]

\author{Yiu Hong Ng and Jim McCann}

\address{Atomic and Molecular Physics Group 
\ftnote{1}{Research URL: 
http://www.dur.ac.uk/$\sim $dph0www1},
Department of Physics, University of Durham,
Durham DH1 3LE, UK.}

\begin{abstract}
Total cross sections for electron capture are calculated 
for collisions of fast protons and $\alpha$--particles 
with atomic hydrogen.
The distorted-wave impulse approximation is applied over 
the energy range 10--1500 keV/u. 
State-selective results are given for the 
$1s$, $2s$ and $2p$ levels. Both the  
{\sl post} and {\sl prior} forms of the model are
calculated and compared with results from other 
theories and experimental measurements. 
In general the model performs very well in comparison with experiment 
over this energy range though discrepancies arise at lower
energies.

\end{abstract}

%
%

\section{Introduction} 
\renewcommand{\theequation}{\arabic{equation}} 

The study of the interaction between atomic hydrogen and fast ions has 
an important role in the physics of
fusion plasmas (Gilbody 1995).
It is significant also in other areas of research such as 
the study of stellar atmospheres (Janev \etal 1987) and the 
physics of proton aurorae (Van Zyl 1993).
An essential requirement for the study of these complex phenomena 
is the availability of  
accurate data for a wide variety of fundamental 
ion-atom collision processes.
Given that the velocity range is such that 
we can consider the nuclear motion as classical and solve
the time-dependent equation of motion for the electrons,
one might suppose that simulations of these processes would be rather simple.
However this is not the case, and the theory behind this problem remains
an area of active investigation (Bransden and McDowell 1993). While
different strategies have been applied 
to problems of this nature,
the most common approach relies on  
basis function expansions (Fritsch and Lin 1991, 
Kuang and Lin 1996, McLaughlin \etal 1997).
This and other methods,  including the 
the direct solution of the partial-differential equation, 
have been reviewed comprehensively by Bransden and McDowell (1993). 

While the basis function method has a well-founded popularity, it
does have some drawbacks. 
Firstly, it can be computationally 
expensive when excited-state coupling is involved. 
Moreover the choice of basis set is problematic;  one cannot always rely 
on the use of larger basis sets to produce better data.
Kuang and Lin (1996) indicate some of the   
spurious effects that can arise if the size and type of a basis 
set are not appropriate.
In spite of these caveats, it is fair to say that recent 
large-scale calculations of this type have given excellent 
agreement with experimental data over the appropriate 
energy range (McLaughlin \etal 1997).
There exist limitations to the method because it is primarily designed 
for strong diabatic coupling between bound states which are
nondegenerate. It tends to 
work less well for higher energies at which ionization (continuum coupling) is
prevalent, and when mixing with nearly degenerate states takes place. 
While significant advances have been made in 
the use of pseudostates to simulate such coupling 
(Slim and Ermolaev 1994, McLaughlin \etal 1997)
the implementation of the method
requires access to high-performance computers. An innovation 
marrying  bound and continuum states has been developed by
Brown and Crothers (1994, 1996a, 1996b)
and includes continuum coupling explicitly in the 
wavefunctions. This is known as the
symmetrized variational continuum distorted-wave ({\sc SVCDW}) method and it
gives excellent agreement with experimental data for electron capture,
excitation and ionization  over the intermediate energy region.

Our paper is concerned with fundamental 
charge exchange processes, namely those occurring between hydrogen
atoms
and fast protons or $\alpha$--particles.
Charge exchange poses theoretical difficulties in that 
one requires an accurate description of the electron dynamics around both
the target and projectile. 
Often a large two-centre basis function expansion is the prescription 
for achieving this aim, or a restricted 
optimized expansion (Brown and Crothers 1994). Considering alternatives to
these
approaches, the fact that charge exchange rapidly decreases in importance 
as the energy is increased  means that one can often use perturbation
theory at high energies to obtain reliable predictions (Belki\'c \etal 1979).
This also has the advantage in being
easy to calculate even with modest computational resources. However,
refinements of perturbation theory, including second-Born type corrections,
are much more difficult to compute and often do not improve the
agreement with experiment (Bransden and McDowell 1993).
This paper investigates the effectiveness of extending one such
{\sl high-energy} approximation into
an {\sl intermediate} energy regime.  
Here we understand the term {\sl intermediate energy} to refer 
to the range 10-200 keV/u, and this is the range of greatest 
interest for beam injection techniques used in fusion science 
(Gilbody 1995). 
The method we adopt, the distorted-wave impulse approximation
({\sc DWIA}), though based upon perturbation theory can account 
for high-order potential scattering from both centres. 
However
the theory is not of the close-coupling type and thus 
in its present form 
cannot describe strong intrashell (Stark) mixing or 
backcoupling (McCann 1992). 
 
The {\sc DWIA} was formulated (Miraglia 1982) as an
improvement upon the conventional plane-wave impulse approximation {\sc  (PWIA)}
for electron capture in ion-atom collisions. 
The {\sc PWIA} is designed to work for collisions in which the collision
is fast in terms of the atomic time-scales (projectile or target).
The validity for the impulse approximation
depends upon the collision time being much shorter than some characteristic
orbital period of the electron bound state. This prerequisite comes
from treating the weaker
interacting nucleus as a spectator particle.
If the collision is sufficiently fast on the atomic time-scale of this 
particle then the spectator particle cannot interact strongly with the
electron during the transition. Its role is confined to  
describing the electronic state before or after the supposed three-body
interaction
in terms of binding the electron and providing it
with a momentum distribution in the bound state.
This establishes a loose validity criterion for the approximation 
that the collision time should be shorter than the internal 
interaction time, or in other words, the atomic orbital
period.
In applications the impulse approximation was found to be very effective in 
reproducing a large variety of experimental data for
very fast collisions  (Jakubassa-Amundsen and Amundsen 1980).
However the inherent 
asymmetry of the method leads to {\sl post} and {\sl prior} forms of
the transition amplitude;  in general these do not agree well. The 
two versions of the approximation converge at high energies, but 
strongly diverge at intermediate and low energies.
On the other hand the  {\sc DWIA} seems to be better suited to 
both symmetric and asymmetric collisions. While we find that there is 
still a {\sl post-prior} discrepancy, it is very much 
less than that which arises from using the {\sc PWIA}.
We employ this model to investigate the following collision processes
 \begin{equation}
{\rm H}^++{\rm H}(1s)\rightarrow {\rm H}(1s,2s,2p)+{\rm
H}^+\label{hydrogen}
\end{equation}
and
\begin{equation}
{\rm He}^{2+}+{\rm H}(1s)\rightarrow {\rm He}^+(1s,2s,2p)+{\rm
H}^+.\label{helium}
\end{equation}
These are ideal processes to test any theory because
the atomic structure is trivial. Also the absence of 
screening effects allows us to write continuum
states in terms of Coulomb functions. For this reason the evaluation
of matrix elements is straightforward.

\section{Method}
In this three-body re-arrangement reaction, we have a bare projectile
nucleus of charge $Z_P$ travelling at a velocity $\mbox{\boldmath $v$}$
relative to the target atom which has nuclear charge $Z_T$. The position of 
the electron in question, with respect to the target and projectile
nuclei, is defined to be $\mbox{\boldmath $r$}_T$ and $\mbox{\boldmath
$r$}_P$, respectively. The initial and final atomic energies of the
electron are denoted by $E_T$ and $E_P$, corresponding to the target
and projectile state electron wavefunctions
$\Phi_{i}$ and $\Phi_{f}$. The momentum-space representations of these
states are labelled as  
$\tilde{\Phi}_{i}$ and $\tilde{\Phi}_{f}$ in this paper.

The definitions of {\sl post} and {\sl prior} can lead to confusion so
let us clarify their meanings. 
We use the standard convention that {\sl post} refers to using the 
impulse approximation in the initial channel for the interaction between
the projectile ion and target atom. In order to calculate the
{\sl prior} form, a separate program must be written.  
\begin{eqnarray}
{\rm DWIA ({\sl post})} \ \ \ \  P({\text{\sc PWIA)}} + (e^-,T)
 \rightarrow (P,e^-) + T({\text{\sc CDW)}} \\
{\rm DWIA ({\sl prior})} \ \ \ \  P({\text{\sc CDW)}} + (e^-,T)
 \rightarrow (P,e^-) + T({\text{\sc PWIA)}} 
\end{eqnarray}
In the {\sl prior} form, the {\sc DWIA}
transition amplitude $a_{fi}^{(-)}($\mbox{\boldmath $b$}$)$ for a given 
impact parameter $\mbox{\boldmath $b$}$  consists
of a continuum
distorted-wave $\xi_{\sc CDW}^{(+)}$ in the entrance channel and a
distorted-wave $\chi_{\sc
IA}^{(-)}$ from the {\sc PWIA} in
the exit channel: 
\begin{equation}
a_{fi}^{(-)}($\mbox{\boldmath $b$}$) =
-i\int\limits^{\infty}_{-\infty}{\rm
d}t\;\langle\chi_{\sc
IA}^{(-)}\; |\; H-i\partial_t\; |\; \xi_{\sc
CDW}^{(+)}\rangle
\end{equation}
The expression in  momentum-space (McCann 1992) 
in the {\sl prior} form is: 
\begin{eqnarray}
\tilde{a}_{fi}^{(-)}({\mbox {\boldmath
$\eta$}})=(2\pi)^{-5/2}v^{-1}iN(a_P)\int {\rm d}{\mbox
{\boldmath $k$}}\; N(a_T)\; \tilde{\Phi}_f^*({\mbox {\boldmath
$k$}}-{\mbox {\boldmath $v$}})\; {\mbox {\boldmath $c$}}\cdot {\mbox
{\boldmath $d$}}\label{aif}\\ 
{\mbox {\boldmath $c$}}=\int {\rm d}{\mbox
{\boldmath $r$}_T}\;  \exp(i{\mbox {\boldmath $\mu\cdot r$}_T})\; 
\;_1F_1\left
[ia_T;1;i(kr_T+{\mbox {\boldmath $k\cdot r$}_T})\right ]
\nabla_{r_T}\Phi_i({\mbox {\boldmath $r$}_T})\label{c}\\
{\mbox {\boldmath $d$}}=\int {\rm d}{\mbox {\boldmath $r$}_P}\; 
\exp(i{\mbox {\boldmath $\omega\cdot
r$}_P})\;\nabla_{r_P}\;_1F_1{\big[}ia_P;1;i(vr_P+{\mbox {\boldmath $v\cdot
r$}_P}){\big]}\label{d} 
\end{eqnarray}
with the momentum transfer vector ${\mbox {\boldmath
$\mu$}}=-{\mbox {\boldmath
$\eta$}}+(\Delta\!E/v^2-{\text{$ 1 \over 2$}})
{\mbox {\boldmath $v$}}$, \;$a_T=Z_T/k$, $\;
a_P=Z_P/v$ and ${\mbox
{\boldmath $\omega$}}=-{\mbox {\boldmath $\mu$}}-{\mbox {\boldmath $k$}}$,
the electron momentum being {\mbox {\boldmath $k$}}. The 
transverse  impulse is denoted by {\mbox
{\boldmath $\eta$}}.
The normalization factor $N(x)$ is defined by $N(x)=\exp(\pi
x/2)\;\Gamma(1-ix)$, and $\Delta\!E=E_P-E_T$. 
The {\sl post} form of the transition amplitude is:
\begin{eqnarray}
\tilde{a}_{fi}^{(+)}({\mbox {\boldmath
$\eta$}})=(2\pi)^{-5/2}v^{-1}iN(a_T^\prime)\int {\rm d}{\mbox
{\boldmath $k$}}\; N(a_P^\prime)\; \tilde{\Phi}_i({\mbox {\boldmath
$k$}}+{\mbox {\boldmath $v$}})\; {\mbox {\boldmath $c^\prime$}}\cdot
{\mbox
{\boldmath $d^\prime$}}\label{aifpost}\\ 
{\mbox {\boldmath $c^\prime$}}=\int {\rm d}{\mbox
{\boldmath $r$}_P}\;  \exp(i{\mbox {\boldmath $\mu^\prime\cdot r$}_P})\; 
\;_1F_1\left
[ia_P^\prime;1;i(kr_P-{\mbox {\boldmath $k\cdot r$}_P})\right ]
\nabla_{r_P}\Phi_f^*({\mbox {\boldmath $r$}_P})\label{cprime}\\
{\mbox {\boldmath $d^\prime$}}=\int {\rm d}{\mbox {\boldmath $r$}_T}\; 
\exp(i{\mbox {\boldmath $\omega^\prime\cdot
r$}_T})\;\nabla_{r_T}\;_1F_1{\big[}ia_T^\prime;1;i(vr_T+{\mbox {\boldmath
$v\cdot
r$}_T}){\big]}\label{dprime} 
\end{eqnarray}
Here the momentum transfer vector ${\mbox {\boldmath
$\mu^\prime$}}={\mbox {\boldmath
$\eta$}}-(\Delta\!E/v^2+{\text{$ 1 \over 2$}})
{\mbox {\boldmath $v$}}$, \;$a_T^\prime =Z_T/v$, $\;
a_P^\prime =Z_P/k$ and ${\mbox
{\boldmath $\omega^\prime$}}=-{\mbox {\boldmath $\mu^\prime$}}+{\mbox
{\boldmath $k$}}$.

For the purposes of our calculation we have considered transitions 
between excited target states and the projectile ground state. 
One can always relate the data obtained to the inverse  
reaction using the principle of detailed balance and time-reversal
symmetry, {\it i.e.} cross
sections for the processes (\ref{hydrogen}) and (\ref{helium}) can be
elicited by looking at their time-reversed counterparts:
\begin{equation} {\rm H}^++{\rm H}(1s,2s,2p)\rightarrow {\rm H}(1s)+{\rm
H}^+\label{hydrogen2} 
\end{equation} 
and 
\begin{equation} {\rm H}^++{\rm
He}^+(1s,2s,2p)\rightarrow {\rm He}^{2+}+{\rm H}(1s).\label{helium2}
\end{equation} 
In our calculations we
have therefore explicitly considered electron capture {\sl into} the $1s$
state from the
$1s$, $2s$ and $2p_{x,y,z}$ states. From now on, we will discuss
calculations in terms of the {\sl prior} form only, for the sake of
convenience.

\subsection{$2s$-$1s$ transition}
Previous work using the {\sc DWIA} (Gravielle and Miraglia
1988)
dealt with electron capture from
the K shell, specifically $1s$-$1s$ electron transfer. It was noted 
(McCann 1992) that the
evaluation of the azimuthal integral, with the polar axis lying along the
momentum transfer vector $\mbox{\boldmath $\mu$}$, generates an Appell
function $F_1$ (Appell and Kamp\'e de Feriet 1926) which is a
hypergeometric function of two variables. Its many linear
transformations and analytic continuations (Olsson 1964) can
be used to evaluate this function.
Nevertheless it remains a major obstacle to efficient and 
accurate computation of the capture cross section.

For the case of $2s$-$1s$ electron transfer, it can easily be shown using
parametric differentiation that
the spatial integrals, (\ref{c}) and ({\ref{d}), over ${\mbox {\boldmath
$r$}_T}$ and ${\mbox {\boldmath $r$}_P}$ simplify in a manner very similar to
that found in the $1s$-$1s$ symmetric transfer calculation (McCann 1992).
Using spherical  coordinates for the 
${\mbox {\boldmath $k$}}$--integral in (\ref{aif}), we denote the angles 
between ${\mbox{\boldmath $\mu $}}$ and ${\mbox{\boldmath
$k$}}$, and that between ${\mbox{\boldmath $\mu $}}$ and ${\mbox{\boldmath  
$v$}}$ by $\theta_{\mu k}$ and $\theta_{\mu v}$. Taking the angle
between these two planes as $\phi$, we then encounter the following
term:
\begin{eqnarray}
\fl\int\limits^{2\pi}_{0}{\rm d}\phi\;
(A+B\cos\phi)^{-2}(C+D\cos\phi)^{-1-ia_P}=2\pi(A+B)^{-2}(C+D)^{-1-ia_P}
\nonumber\\ \;\;\;\;\;\;\;\;\;\;\;\;\;\;\;\;\;\;\;\;\;\times
F_1({\text{$1\over
2$}};2,1+ia_P;1;x_1,x_2) 
\end{eqnarray} with $x_1=2B/(A+B)$ and $x_2=2D/(C+D)$. The quantities $A$,
$B$, $C$ and $D$ have the definitions:
\begin{eqnarray}
A=Z_P^2+v^2+k^2-2vk\cos{\theta_{\mu k}}\cos{\theta_{\mu v}}\nonumber\\
B=2vk\sin{\theta_{\mu k}}\sin{\theta_{\mu v}}\nonumber\\
C=-\Delta\!E+v^2/2-vk\cos{\theta_{\mu k}}\cos{\theta_{\mu v}}\nonumber\\
D=vk\sin{\theta_{\mu k}}\sin{\theta_{\mu v}}\label{quant}
\end{eqnarray}

\subsection{$2p$-$1s$ transition} 
The lack of spherical symmetry for the
$2p$ sub-states means that the choice of the axis of quantization is
important
(Coleman and Trelease 1968).  Using the momentum transfer vector ${\mbox
{\boldmath $\mu$}}$ as the polar axis simplifies calculations and 
the change to the laboratory frame of reference  requires
a simple rotation transformation.

A different line of approach takes advantage of parametric differentiation.
We can consider the spherical harmonics in their real form:
\begin{eqnarray}
\Phi_{2p_j}( {\mbox{\boldmath $r$}_T} )
={\text{$1\over 4$}}Z_T^{5/2}(2\pi)^{-1/2}\exp(-\lambda_Tr_T)\;r_{Tj}
\end{eqnarray}
where $\lambda_T= {\text{$1\over 2$}}Z_T$
and $j\in\{x, y, z\}$.
In evaluating (\ref{c}), integration by parts gives:
\begin{eqnarray}
{\mbox{\boldmath $c$}}={\mbox{\boldmath
$I_1$}}+{\mbox{\boldmath
$I_2$}}\\
{\mbox{\boldmath $I_1$}}=-\int{\rm
d}{\mbox{\boldmath $r$}_T}\;
\exp(i{\mbox {\boldmath $\mu\cdot r$}_T})\; \Phi_i({\mbox{\boldmath
$r$}_T})\; \nabla_{r_T}\;_1F_1{\big [}ia_T;1;i(kr_T+{\mbox {\boldmath
$k\cdot
r$}_T}){\big ]}\\
{\mbox{\boldmath $I_2$}}=-i{\mbox{\boldmath
$\mu$}}\int{\rm
d}{\mbox{\boldmath $r$}_T}\;
\exp(i{\mbox {\boldmath $\mu\cdot r$}_T})\;_1F_1{\big 
[}ia_T;1;i(kr_T+{\mbox
{\boldmath $k\cdot 
r$}_T}){\big ]}\label{i2}
\end{eqnarray}
The integral ${\mbox{\boldmath $I_1$}}$ is further simplified by writing
it as:
\begin{eqnarray}
\fl {\mbox{\boldmath $I_1$}}=-q{\partial
\over{\partial
{\mbox{\boldmath
$q$}}}}\int{\rm d}{\mbox{\boldmath $r$}_T}\; {\exp(i{\mbox {\boldmath
$\mu\cdot r$}_T}) \over {r_T}}\Phi_i({\mbox{\boldmath
$r$}_T})_1F_1{\big [}ia_T;1;i(qr_T+{\mbox
{\boldmath $q\cdot
r$}_T}){\big ]}{\bigg\arrowvert}_{{\mbox{\boldmath $q$}}={\mbox{\boldmath
$k$}}}\nonumber\\
\lo={\text{$1\over 4$}}{Z_T}^{5/2}(2\pi)^{-1/2}iq{\partial
\over{\partial
{\mbox{\boldmath
$q$}}}}{\partial \over{\partial \mu_j}}\int{\rm d}{\mbox{\boldmath
$r_T$}}\; {\exp(-\lambda_Tr_T+i{\mbox {\boldmath
$\mu\cdot r_T$}}) \over {r_T}}\nonumber\\
\times_1F_1{\big [}ia_T;1;i(qr_T+{\mbox
{\boldmath $q\cdot
r_T$}}){\big ]}{\bigg\arrowvert}_{{\mbox{\boldmath $q$}}={\mbox{\boldmath
$k$}}}\nonumber\\
\lo={\text{$1\over 4$}}{Z_T}^{5/2}(2\pi)^{-1/2}iq{\partial
\over{\partial
{\mbox{\boldmath
$q$}}}}{\partial \over{\partial \mu_j}}{\bigg [}{2\pi\alpha^{ia_T-1}\over
{(\alpha+\beta)^{ia_T}}}{\bigg ]}{\bigg\arrowvert}_{{\mbox{\boldmath
$q$}}={\mbox{\boldmath
$k$}}} 
\end{eqnarray}
where $\alpha={1\over 2}({\lambda_T}^2+\mu^2)$, $\;\beta={\mbox {\boldmath
$\mu\cdot
q$}}-i\lambda_T\;q$, and $\partial/\partial\mu_j$ calls for partial
differentiation with respect to the $j$ component of ${\mbox{\boldmath
$\mu$}}$, $j\in\{ x,y,z\}$.
The momentum transfer vector ${\mbox{\boldmath
$\mu$}}$ consists of two mutually orthogonal components, with one running
parallel to the direction of travel $\hat{\mbox{\boldmath
$v$}}$ of the incoming projectile which we can take as the $z$-axis. The
transverse component is $-{\mbox{\boldmath
$\eta$}}$ and this lies within the $xy$-plane. It is then convenient
to let ${\mbox{\boldmath
$\eta$}}$ lie along the $x(y)$-axis with
the result that capture from the $2p_{y(x)}$
state is parity forbidden.

Employing the same method, the expression (\ref{i2})
 can be shown to be given by:
\begin{equation}
{\mbox{\boldmath $I_2$}}={\text{$1\over
4$}}{Z_T}^{5/2}(2\pi)^{-1/2}{\mbox{\boldmath $\mu$}}{\partial
\over{\partial
\lambda_T}}{\partial \over{\partial \mu_j}}{\bigg
[}{2\pi\alpha^{ia_T-1}\over
{(\alpha+\beta)^{ia_T}}}{\bigg ]}
\end{equation}
\begin{table}[t]
\caption{Cross sections $\sigma_{nl}$ ($10^{-17} {\rm cm}^2$) of
electron capture for
proton-hydrogen
collisions at energy E keV: $1s$ to $nl$ transition. The integer in
parenthesis indicates 
the power of ten by which the number has to be multiplied.} 
\footnotesize\rm
\begin{indented}
\item[]\begin{tabular}{@{}llllllll} 
\br 
E(keV) && 125 & 250 & 500
 & 750 & 1000 & 1500\\ \mr $\sigma_{1s}$ & DWIA & 2.45(-1) &
1.35(-2)
& 4.83(-4)  & 5.97(-5) & 1.30(-5) & 1.43(-6)\\ &
CDW
& 2.70(-1) & 1.38(-2) & 4.83(-4)  & 5.96(-5) &
1.29(-5) & 1.43(-6)\\\mr 
$\sigma_{2s}$
&DWIA$_{post}$&4.68(-2)&2.19(-3)&6.98(-5)&8.23(-6)&1.74(-6) 
&1.87(-7)\\
&DWIA$_{prior}$&4.50(-2)&2.20(-3)&7.02(-5)&8.28(-6)&1.75(-6)
&1.88(-7)\\
& CDW & 5.05(-2)&2.26(-3) & 7.13(-5)
&8.43(-6)&1.79(-6)&1.93(-7)\\\mr
$\sigma_{2p}$
&DWIA$_{post}$& 9.89(-3)& 2.65(-4)&4.93(-6)&4.42(-7)&7.78(-8)&
6.54(-9)\\
&DWIA$_{prior}$&1.29(-2)&3.76(-4)&7.38(-6)&6.60(-7)&1.15(-7)
&9.67(-9)\\
&CDW&1.54(-2)&4.83(-4)&1.01(-5)&9.53(-7)&1.74(-7)  &1.58(-8)\\ \br
\end{tabular} \end{indented} \end{table}

For electron capture from the $2p$ states, the azimuthal integral of
(\ref{aif}) has the form:
\begin{equation}
J=\int\limits^{2\pi}_{0}{\rm d}\phi\; (A^\prime +B^\prime\cos\phi)^{-2}\;
(C^\prime +D^\prime\cos\phi)^{-1-ia_P}\; (E^\prime +F^\prime\cos\phi)
\label{J}
\end{equation}
Like those of (\ref{quant}), the quantities $A^\prime$,
$B^\prime$, $C^\prime$, $D^\prime$,
$E^\prime$ and $F^\prime$ depend
upon ${\mbox{\boldmath $\eta$}}$, ${\mbox{\boldmath $k$}}$ and the polar
angle between ${\mbox{\boldmath $k$}}$ and ${\mbox{\boldmath $\mu$}}$, but
are much more complicated; we do not quote the details.
 
Parametric differentiation transforms (\ref{J}) into an expression
containing
derivatives of the more familiar Appell functions $F_1$.
\begin{eqnarray}
\fl J=-{\big (}E^\prime{\partial\over\partial
A^\prime}+F^\prime{\partial\over\partial
B^\prime}{\big )}\int\limits^{2\pi}_{0}{\rm d}\phi\;
(A^\prime +B^\prime\cos\phi)^{-1}\;
(C^\prime +D^\prime\cos\phi)^{-1-ia_P}\nonumber\\
\lo=-{\big (}E^\prime{\partial\over\partial
A^\prime}+F^\prime{\partial\over\partial
B^\prime}{\big )}{\big \{
}2\pi(A^\prime +B^\prime )^{-1}(C^\prime +D^\prime)^{-1-ia_P}\nonumber\\
\times\; F_1({\text{$1 \over 2$}};1,1+ia_P;1;x_1,x_2){\big
\} }
\end{eqnarray}
Since
\begin{equation}
\partial_{x_1}F_1(a;b_1,b_2;c;x_1,x_2)={ab_1\over
c}F_1(a+1;b_1+1,b_2;c+1;x_1,x_2)
\end{equation}
we therefore have:
\begin{eqnarray}
\fl J=-2\pi(A^\prime +B^\prime )^{-3}(C^\prime +D^\prime )^{-1-ia_P}{\big
\{
}F_1({\text{$3\over 2$}};2,1+ia_P;2;x_1,x_2)\;(A^\prime
F^\prime -B^\prime E^\prime )\nonumber\\
\lo-F_1({\text{$1\over 2$}};1,1+ia_P;1;x_1,x_2)\;(A^\prime
+B^\prime )(E^\prime +F^\prime ){\big
\}
}
\end{eqnarray}
As with $1s$-$1s$ electron capture, the corresponding $2s$-$1s$ and
$2p$-$1s$ transition amplitudes possess two singularities, at $C\pm D=0$
and $C^\prime \pm D^\prime =0$ respectively, which are integrable.

\subsection{Symmetric models} 

In the {\sc DWIA} we inevitably have a {\sl post}-{\sl prior} asymmetry.
If 
there is an inherent asymmetry in the reaction, for example
$Z_T \gg Z_P $, the choice between {\sl post} and {\sl prior} models can
be
argued in favour of one or the other.
A detailed discussion of this question is given in the next section. 
A merit of the {\sc CDW} theory is the intrinsic {\sl post}-{\sl prior}
equivalence.
Symmetry can be artificially introduced in our model by  averaging the
{\sl post} and {\sl prior} amplitudes. The resulting symmetrized {\sc
DWIA} is given by:
\begin{equation}
\tilde{a}_{fi}^S({\mbox {\boldmath $\eta$}})
\equiv
{\text{$ 1 \over 2$}} \left[ \tilde{a}_{fi}^{(-)}({\mbox {\boldmath $\eta$}})
+ \tilde{a}_{fi}^{(+)}({\mbox {\boldmath $\eta$}}) \right]
\label{cdwia}
\end{equation}
Miraglia (1982) had also proposed making the impulse approximation 
in both entry and exit channels (generalized 
impulse approximation {\sc GIA}). While it is much more difficult to
calculate the {\sc GIA}, it is not clear whether 
this {\sc GIA} model will have a larger range of validity 
than the {\sc DWIA}, and thus whether it offers a substantial gain in 
accuracy.

\section{Results and discussion}
\begin{table}[t] 
\caption{Cross sections $\sigma_{nl}$ ($10^{-17} {\rm cm}^2$) of
electron capture for helium nucleus-hydrogen
collisions at energy E keV: $1s$ to $nl$ transition. The integer in
parenthesis
indicates 
the power of ten by which the number has to be multiplied.} 
\begin{indented}
\item[]\begin{tabular}{@{}llllllll} 
\br 
E(keV) && 10 & 25 & 60
 & 100 & 150 & 200\\ \mr 
$\sigma_{1s}$&DWIA$_{post}$&2.76(0)&2.40(0)&9.02(-1)&4.87(-1)&2.42(-1)
&1.25(-1)\\
&DWIA$_{prior}$&3.66(0)&4.22(0)&2.10(0)&8.58(-1)&3.24(-1)&1.41(-1)\\
&CDW&2.78(1)&3.07(0)&2.73(0)&1.10(0)&3.94(-1)&1.64(-1)\\
&SVCDW&2.92(-1)&9.93(-1)&1.70(0)&7.67(-1)&2.86(-1)&1.21(-1)\\
\mr
$\sigma_{2s}$&DWIA$_{post}$&2.07(1)&4.81(0)&1.30(0)&4.62(-1)&2.54(-1)& 
5.42(-2)\\
&DWIA$_{prior}$&1.77(1)&5.86(0)&1.33(0)&4.47(-1)&1.38(-1)&5.11(-2)\\
&CDW&1.68(2)&2.78(1)&3.53(0)&7.84(-1)&1.94(-1)&6.47(-2)\\
&SVCDW&1.58(1)&9.41(0)&1.90(0)&4.72(-1)&1.25(-1)&4.44(-2)\\
\mr
$\sigma_{2p}$&DWIA$_{post}$&1.05(2)&1.46(1)&2.26(0)&6.21(-1)&1.53(-1) 
&4.52(-2)\\
&DWIA$_{prior}$&1.32(2)&3.54(1)&5.79(0)&1.02(0)&1.93(-1)&5.20(-2)\\
&CDW&1.00(3)&1.10(2)&8.01(0)&1.23(0)&2.24(-1)&6.00(-2)\\
&SVCDW&6.07(1)&2.46(1)&2.95(0)&5.32(-1)&1.12(-1)&3.18(-2)\\
\mr 
\end{tabular}
\end{indented} 
\end{table}

We present
results for capture cross sections as a function of energy 
and compare this data with experiment and other models.
For the most part we use 
logarithmic graphs, but in addition we have compiled 
a small sample of the data in tabular form for reference purposes
(tables 1 and 2). This helps to form detailed comparison of
the {\sl post} and {\sl prior} forms of the {\sc DWIA} results which would
not
be apparent from the graphs. 
We have used the {\sc CDW} model as a benchmark (Belki\'c \etal 1979) 
for the results. This model reproduces electron capture data very well
over the high-energy range but suffers from a lack of unitarity at
lower energies and invariably produces gross overestimates for 
capture at intermediate energies. 

In figure 1 the {\sc DWIA} results for 
electron capture cross sections of
the $1s$-$2s$ transition in reaction (\ref{hydrogen}) are shown.
The corresponding data of Brown and Crothers (1996a) have not been
presented,
although they follow
the experimental results very closely, even down to 10 keV/u.
Instead our results are compared with
experimental data and the continuum distorted-wave ({\sc CDW}) results
over the intermediate energy range: 20--250 keV/u. 
At the upper end of the energy regime we note that both {\sc DWIA}
curves merge with the {\sc CDW} results. However this 
convergence is not uniform. If we refer to table 1 for
energies beyond the scope of the graph, apart from
noting the sharp fall in the
size of the cross section ($\sigma_{2s}$), it is clear that 
at very high energies the 
{\sl post} and {\sl prior} results converge but begin to depart from the
{\sc CDW} data. This can be understood from 
noting the importance of second-order effects at these energies.  
The {\sc CDW} theory can partly but not fully account for these 
effects, and the {\sc DWIA} data are more reliable estimates in this
case. Nonetheless the {\sc CDW} results are more than adequate for
good estimates of cross sections at high energies.

The intermediate energy range covered by figure 1 indicates the
disparity of these models more clearly. The {\sc CDW} results
continue rising sharply as is well-known (Belki\'c \etal 1979). 
This is typical 
of the difficulties in applying high-energy perturbation 
theory over the intermediate energy range. We note that {\sc DWIA} gives a
slight improvement over {\sc CDW} below around 50 keV/u, although its
accord with experiment remains poor at lower energies. 
Furthermore the {\sl post} and {\sl prior} {\sc DWIA}
models show large differences below 50 keV/u. In common with
the {\sc PWIA}, the inherent {\sl post}-{\sl prior}
asymmetry is expected to be greatest at the lowest velocities. As
a general rule, the
discrepancy is worse for {\sc PWIA} than for {\sc DWIA} (Ng and McCann 1997).

On theoretical grounds we can propose a 
prescription for the energy range of validity of the {\sc DWIA} and for
the preferred form ({\sl post} or {\sl prior}) of the theory in this
particular
reaction. Using the peaking
criterion (McCann
1992), the orbital velocity of the electron in the final state is smaller 
than that in the initial state, and so one should apply {\sc CDW}    
refinements in the exit channel and allow the {\sc PWIA} to take account
of momentum spread in the entrance channel. This implies that the 
{\sl post} form should be preferred. Next, one can use the impulse
hypothesis (Gravielle and Miraglia 1988) 
to establish the energy limit - given the largest
orbital velocity is 1 a.u. the desired stipulation would be that $E \gg
25$ keV/u.
This inequality  is consistent with the experimental data 
in figure 1.
Since we wish to know the practical limits of our model, the important
question here is: how strong is the inequality? In this paper we
aim to answer this question by a mixture of comparisons with
experiment and other theories. 

We have also computed, for the sake of completeness, several
results using the symmetrized
{\sc DWIA} (equation \ref{cdwia}). The results are shown in figure 1 and give
the best agreement of all the calculations. Nonetheless this averaging
procedure seems rather artificial to us, and the presence of
a large {\sl post}-{\sl prior} discrepancy at the lower energies indicates
that 
neither the {\sl post} nor the {\sl prior} form is very satisfactory
around 25 keV/u. 
On the whole the theory of Brown and Crothers (1996a) still appears to
be the most satisfactory description of the process both in 
physical and quantitative terms.

Figure 2 shows capture cross sections for $1s$-$2p$ transition. Below 40
keV/u, {\sc DWIA} again gives slightly better results than {\sc CDW}. In
the
high-energy range the results do not converge. Both sets of
{\sc DWIA} results differ and they in turn are lower than the 
{\sc CDW} results by a fairly constant ratio (table 2).
Again the {\sl post} {\sc DWIA} is expected to be the best physical 
model for this process and this is confirmed by the better agreement
with the {\sc SVCDW } of Brown and Crothers (1996a) over
the energy range 20--100 keV/u. It has been noted that the experimental
results all have systematic uncertainties with regard to the normalization of
the data. However these uncertainties are not large and a
series of experiments has been carried out 
on this reaction process using independent 
estimates of the absolute magnitudes of the cross sections. It
is very likely the experimental data in the figure are reliable both in
terms of the energy dependence and absolute values.
In any case we have not attempted to renormalize the data to theoretical
predictions. 

The process (\ref{hydrogen}) is dominated by $1s$ and $2s$ 
transitions above 100 keV/u. For
higher energies we can estimate capture
into higher excited states 
($\sigma_n$) by
the rough approximation based on the $n^{-3}$-distribution of
populations.  The total cross
section therefore becomes $\sigma_{total}=\sigma_1+1.62\sigma_2$ where
$\sigma_1$ and $\sigma_2$ are cross sections of capture into the $n=1$ and
$n=2$ levels respectively. Results for {\sc DWIA} and {\sc CDW}, in the
energy range
125--2500 keV/u (figure 3), are very similar and are in good agreement
with
experimental data.

The proton-hydrogen collision calculations do not contain
much {\sl post-prior discrepancies} due to the nature of the charge
exchange process being more or less symmetric.
Their agreement with experimental data is very satisfactory. 
However the study of reaction (\ref{helium})
which has a strong in-built asymmetry would provide a much
clearer critical assessment of {\sc DWIA}. But first, the nature and
implication of
the asymmetry must be considered with care.

Generally in electron transfers from the ground state to one which is
excited, an
important point to consider is the momentum spread of the electron,
before and after the collision. A sharply
peaked momentum spread could be treated more efficiently by {\sc CDW}
with its peaking approximation. The impulse approximation should then
be applied to the electronic state with the less sharply peaked momentum
distribution. The momentum spread can be
affected by the electron's
distance from the nucleus and also by the magnitude of the nuclear charge,
and is roughly of the order of $\sim Z/n$.
Since the only asymmetry in proton-hydrogen collision comes from the
difference in initial and final states and not from projectile and target 
nuclear charges, its {\sl post-prior} discrepancies are not appreciable.
The same,
however, cannot be said for (\ref{helium}). In the forward reaction, we
would expect an electron attached to the heavier (projectile)
$\alpha$--particle to have quite a broad momentum distribution. Since this
situation arises after
the collision, it would seem reasonable to use the {\sl prior} form of
{\sc DWIA}. 
The {\sl prior} form corresponds to a {\sc PWIA} wavefunction in the exit
channel, making the bare proton of the hydrogen
atom more of a spectator particle than the helium
nucleus. 
It has been argued 
(Gravielle and Miraglia 1991) that the converse should be true, the
argument being based 
upon the idea of associating the impulse approximation with the
stronger
potential ({\sl i.e.} in the entrance channel).
We take the opposite viewpoint as the weight of evidence
presented in this paper lends it sufficient support.
Consequently for the
time-reversed reaction ({\ref{helium2}), we apply inversion 
so that its {\sl post} form becomes favourable. 

Cross sections for selected energies for reaction (2) are tabulated in
table 2, the energy range (7--200 keV/u) being the same as that covered by
figures 4 and 5. These show cross sections of capture into the $2s$ and 
$2p$ states.
Comparison is made with the
symmetrized variational continuum distorted-wave ({\sc SVCDW}) method
(Brown and Crothers 1996b), the conventional {\sc CDW}
theory and experimental results. In the case of the
$2s$ transition {\sc DWIA} fares rather poorly.
While the curves follow the {\sc SVCDW} curve at higher energies and avoids 
the low-energy divergence of the {\sc CDW}, the agreement with experiment
is not good over the important plateau region (figure 4).
The {\sc DWIA} completely fails to describe this feature.
Results for the $2p$ transition (figure 5) on the other hand are less
conclusive. 
The shortcomings of the {\sc DWIA} around 50 keV/u
for the $2s$ results could be 
attributed to the lack of strong coupling between the resonant states 
(H-$1s$, He$^+$-$2s,2p$). Although the {\sc DWIA}
underestimates the $2s$ cross sections (figure 4) and overestimates 
the $2p$ results (figure 5), this can be explained by a second-order
process involving the redistribution of populations: $H(1s)\rightarrow
He^+(2p)\rightarrow He^+(2s)$.
The inclusion of this effect in the {\sc SVCDW}
model brought
good agreement with experiment. Its neglect in the {\sc
DWIA} means that there is a disconcerting gap between theory
and experiment in figure 4, with a fraction of the cross section
that should have gone towards the $2s$ transition ending up in the $2p$
instead.

The total cross section for capture into all states is shown
in figure 6.
The {\sc DWIA} {\sl prior} curve, with the $n^{-3}$ scaling law included,
performs remarkably well against both {\sc SVCDW} and
{\sc CDW}. Major contribution to the total cross section comes from capture
into the $2p$ state and {\sc DWIA} models this quite well in its {\sl
prior} form, yielding good results.
The use of the $1s$, $2s$, $2p$ results to extrapolate for capture to
all states is doubtful, of course. A simple test shows that when
the $n^{-3}$ scaling law is applied to obtain {\sc CDW} results for
$\sigma_n \ \ (n > 2) $, it 
underestimates the true {\sc CDW} $\sigma_n$ in
the energy range 7--200 keV/u.  This has the effect that the 
extrapolation can be in error by as much as 60\% . In view of this, {\sc
DWIA}
results for total cross sections should be considered as underestimates for
capture to excited states and this could explain part of the shortfall 
between theory and experiment that we see in figure 6.
If we were to assume that the underestimation for {\sc CDW} will be of
the
same order of magnitude for {\sc DWIA} at a fixed energy,
appropriate augmentation of the results
would bring them quite close to the experimental
values.

\section{Conclusions}

In conclusion, despite its inherent {\sl post-prior} discrepancies,
{\sc DWIA} can
perform well even at energies as low as 20 keV/u if sufficient care
is taken in choosing whether to use the {\sl post} or {\sl prior} form of
the theory. This is very useful in establishing the energy range of
validity. However the computational effort in evaluating 
{\sc DWIA} is much
greater than either {\sc CDW} or {\sc PWIA} model; hours as opposed to
seconds on one of our RS/6000 workstations. The additional effort brings
its reward in terms of better estimates of cross sections towards the
upper end of the intermediate energy range. This does provide 
very useful data for 
applications. There are awkward features of the calculation 
that make it difficult to compute without careful attention. 
However the method is significantly less expensive to compute than 
large-scale close-coupling methods, though  it
does not give the same accuracy or
reliability over the intermediate energy regime. It is 
difficult to extend the {\sc DWIA} method further.
One might attempt a close-coupling 
approach akin to the {\sc SVCDW} method. We take the view that 
significant improvements are difficult within the {\sc DWIA} framework 
and that instead the {\sc SVCDW}   
approach seems to combine the correct physical and mathematical
features of the charge transfer processes in an elegant manner. This
approach seems the most promising avenue for future work.

\ack

We are grateful to the UK Engineering and Physical Sciences 
Research Council for their support of this work through a
research studentship and provision of computing resources. The EPSRC-funded
Durham/Newcastle Atmol
cluster was used along with the Columbus workstation cluster 
at Rutherford-Appleton Laboratory.
We are grateful to Dr Geoff Brown for providing us with numerical 
data for the {\sc SVCDW} results for $\alpha$--particle collisions,
and for useful discussions on this subject.

\section*{References}
\begin{harvard}
\item[]Appell P and Kamp\'e de Feriet J 1926 {\it Fonctions
Hyperg\'eometriques et Hypersph\'eriques} (Paris:
Gauthier-Villars)
\item[]Barnett C F 1990 {\it Oak Ridge National Laboratory} Report No 6086
(unpublished)
\item[]Belki\'c Dz, Gayet R and Salin A 1979 {\it Phys. Rep.} {\bf 56} 279. 
\item[]Bransden B H and McDowell M R C 1992
      {\it Charge Exchange and the Theory of
       Ion-Atom Collisions (Oxford: Oxford Science Publications)}{}
\item[]Brown G J N and Crothers D S F 1994 {\it J.Phys. B: At. Mol. Opt.
Phys.} {\bf 27} 5309.
\item[]\dash 1996a {\it Phys. Rev. Lett.}
 {\bf 76} 392
2\item[]\dash 1996b {\it J.Phys. B: At. Mol. Opt.
Phys.} {\bf 29} L705
\item[]\'Ciri\'c D, Dijkkamp D, Vlieg E and de
Heer F J
1985 {\it
J.
Phys. B: At. Mol. Opt. Phys.}
{\bf 18} L17
\item[]Coleman J P and Trelease S 1968 {\it Proc. R. Soc.} {\bf 85} 1097
\item[]Fritsch W and Lin C D 1991 {\it Phys. Rep.} {\bf 202} 1
\item[]Gilbody H B and Ryding G 1966 {\it Proc. R. Soc. (London)} {\bf
A291}
438
\item[]Gilbody H B 1995 {\it XIX ICPEAC Invited Papers, 
       British Columbia, Canada,
       Edited by L J Dub\'e, J B A Mitchell, J W McConkey
       and C E Brion
      (AIP Press: New York)} {{\bf 360} 19}
\item[]Gravielle M S and Miraglia J E 1988 {\it Phys. Rev.} A {\bf 38}
5034
\item[]\dash 1991 {\it Phys. Rev.} A {\bf 44}
7299
\item[]Jakubassa-Amundsen D H and Amundsen P 1980 {\it Z. Phys.} A {\bf
297} 203
\item[]Janev R K, Langer W D, Evans Jr. K and Post Jr. D E 1987 {\it 
       Elementary Processes in Hydrogen-Helium Plasmas 
       (Berlin and New York: Springer-Verlag)}{}
\item[]Kuang Y and Lin C D 1996 {\it J. Phys. B: At. Mol. Opt. Phys.} {\bf 29}
1027
\item[]McCann J F 1992 {\it J.Phys. B: At. Mol. Opt. Phys.} {\bf 25}
449
\item[]McLaughlin B M, Winter T G and McCann J F 1997 {\it J. Phys. B: At. 
Mol. Opt. Phys.} {\bf 30} 1043. 
\item[]Miraglia J E (1982) {\it J. Phys. B: At. Mol. Phys.} {\bf 15} 4205. 
\item[]Morgan T J, Stone J and Mayo R 1980 {\it Phys. Rev.} {\bf A22}
1460
\item[]Ng Y H and McCann J F 1997 (in preparation)
\item[]Olsson P O M 1964 {\it J. Math. Phys.} {\bf 5} 420
\item[]Schwab W, Baptista G B, Justiniano E, Schuch R, Vogt H and Weber E
W 1987
{\it J.
Phys. B: At. Mol. Opt. Phys.} {\bf 20} 2825
\item[]Shah M B and Gilbody H B 1978 {\it J.Phys. B: At. Mol. Opt. Phys.}
 {\bf 11} 121
\item[]Slim H A and Ermolaev A M 1994 {\it J.Phys. B: At. Mol. Opt. Phys.}
 {\bf 27} L203
\item[]Van Zyl B 1993 {\it XVIII ICPEAC Invited Papers, Aarhus,
       Denmark, Edited by T Anderson, B Fastrup, 
       F Folkmann, H Knudsen and N Andersen
      (AIP Press: New York)} {{\bf 295} 684}
\item[]Toburen L H, Nakai M Y and Langley R A 1968 {\it Phys. Rev.} {\bf
171} 114
\end{harvard}
\section*{Figure Captions}
\begin{itemize}
\item Figure 1: Cross sections for $ 1s \rightarrow  2s $
capture for process (\ref{hydrogen}): {\sc
DWIA}({\sl prior}) (\full ), {\sc DWIA}({\sl post}) (\dashed ),
symmetrized {\sc DWIA} ($\diamond$), {\sc CDW}
(\dotted) and experimental results by Morgan \etal (1980)
($\bullet $). 

\item Figure 2: Cross sections for $ 1s \rightarrow 2p$
electron capture for process (\ref{hydrogen}): {\sc DWIA}({\sl prior}) (\full ),
{\sc DWIA}({\sl post}) (\dashed ), {\sc CDW} (\dotted) and
experimental results by Barnett (1990) ($\bullet $).

\item Figure 3: Total cross section of electron capture into all states
in collisions between proton and hydrogen: {\sc DWIA} (\full ), {\sc CDW}
(\dotted)
and experimental results by Toburen {\it et al} (1968) ($\bullet $),
Gilbody and Ryding (1966) ($\triangle$) and Schwab \etal (1987)
($\Box $).

\item Figure 4: Cross sections for  $1s \rightarrow 2s$ electron capture
for process (\ref{helium}): {\sc DWIA}({\sl prior})
(\full ),
{\sc DWIA}({\sl post}) (\dashed), {\sc CDW}
(\dotted), {\sc SVCDW} (\chain ) and experimental results by Shah and
Gilbody
(1978) ($\bullet $).

\item Figure 5: Cross sections for $1s \rightarrow 2p$ capture
for process (\ref{hydrogen}): {\sc DWIA}({\sl prior})
(\full ),
{\sc DWIA}({\sl post}) (\dashed),
 {\sc CDW}
(\dotted), {\sc SVCDW} (\chain ) and experimental results by
\'Ciri\'c \etal
(1985) ($\bullet $).

\item Figure 6: Total cross section of electron capture into all states
in
Collisions between helium nuclei and hydrogen: {\sc DWIA}({\sl prior})
(\full ),
{\sc DWIA}({\sl post}) (\dashed ), {\sc CDW}
(\dotted), {\sc SVCDW} (\chain ) and experimental results by Shah and
Gilbody
(1978) ($\bullet $).
\end{itemize}
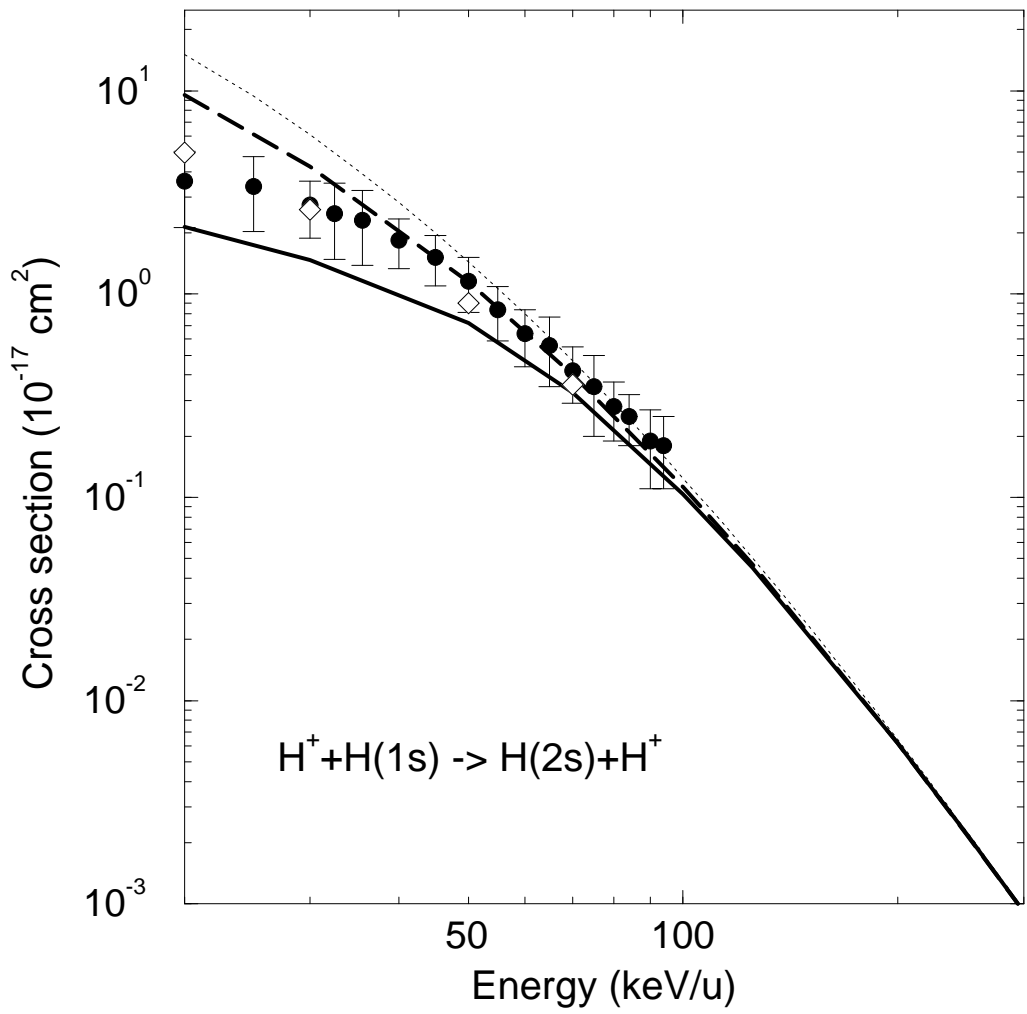
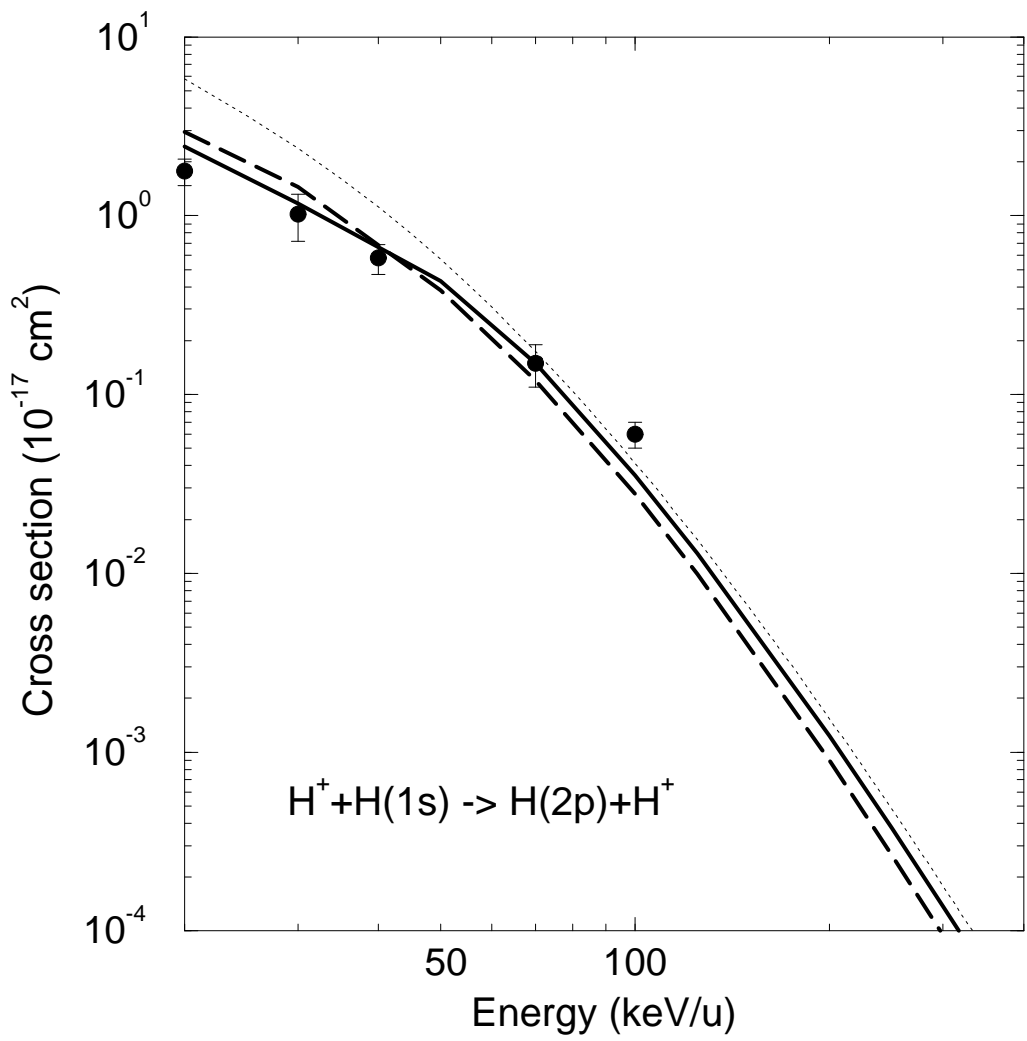
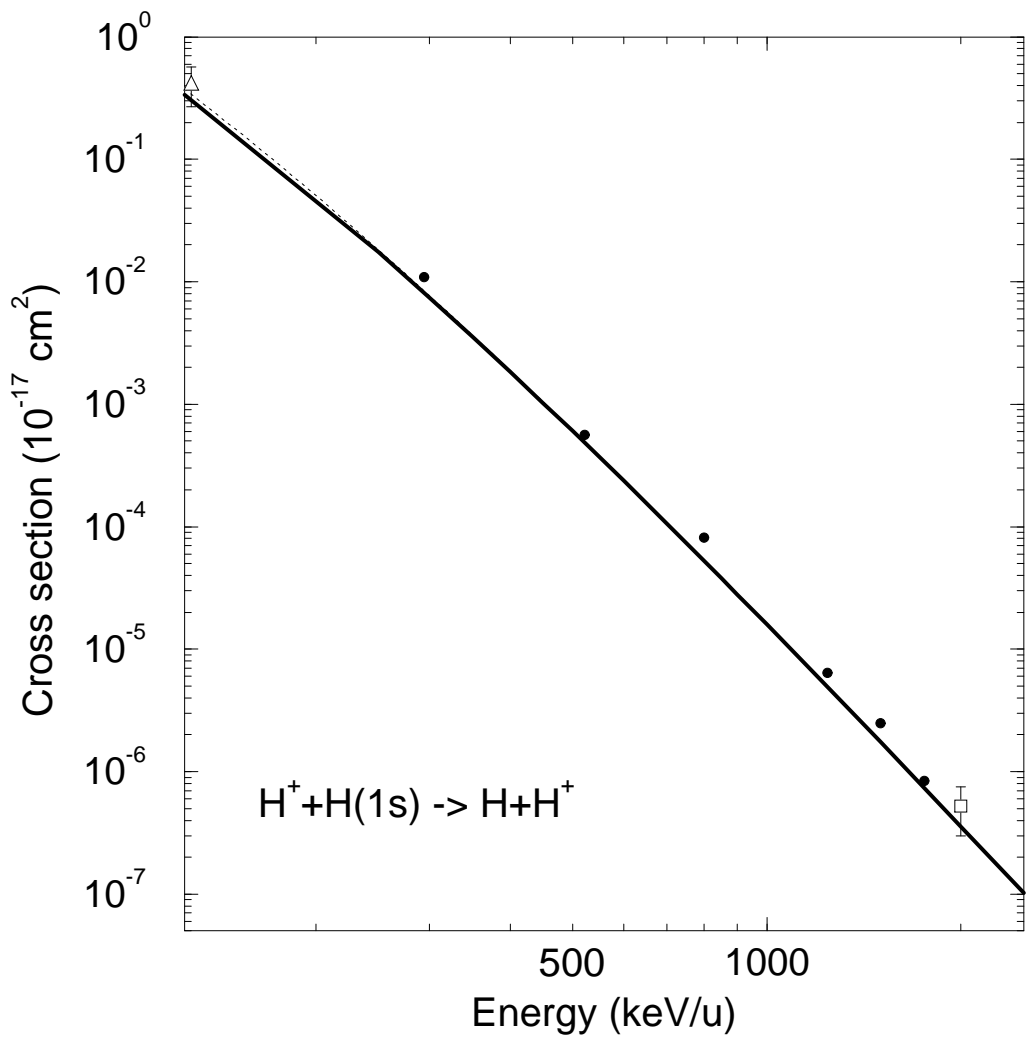
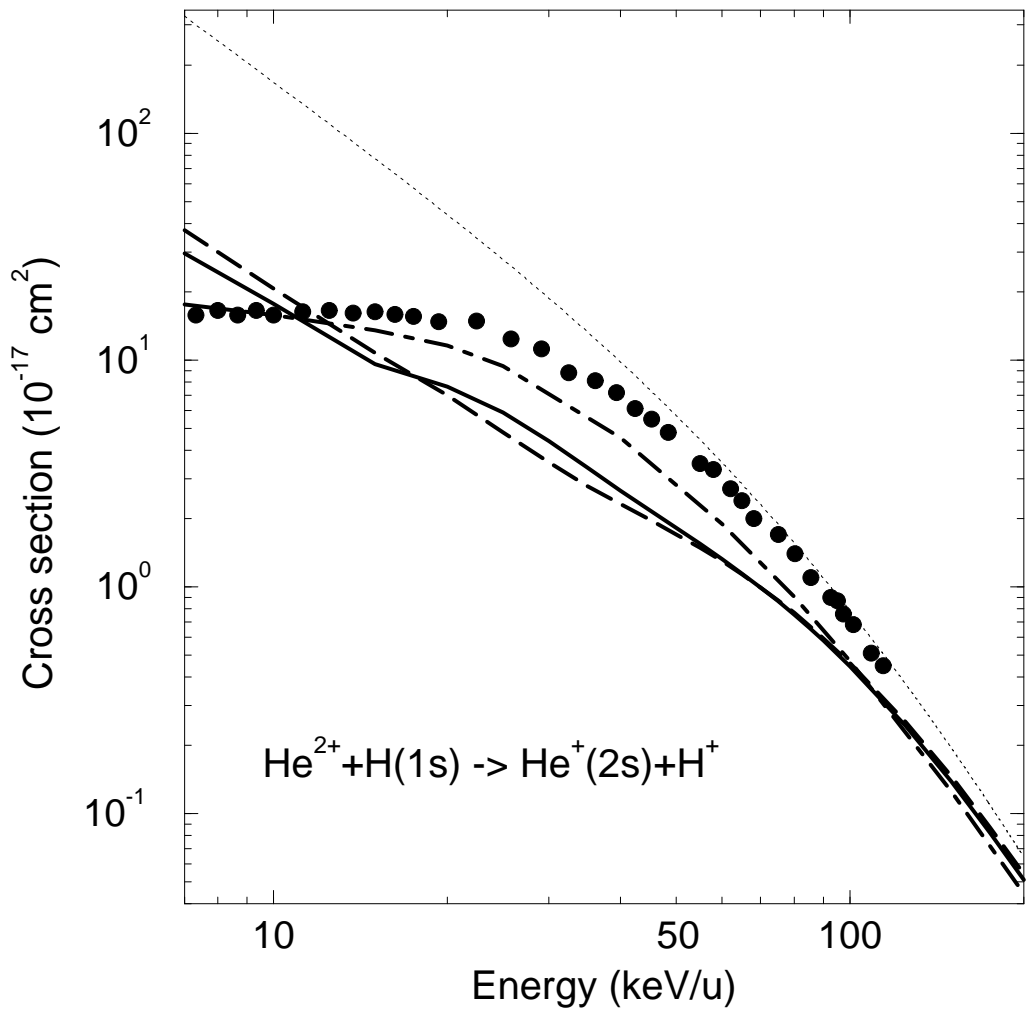
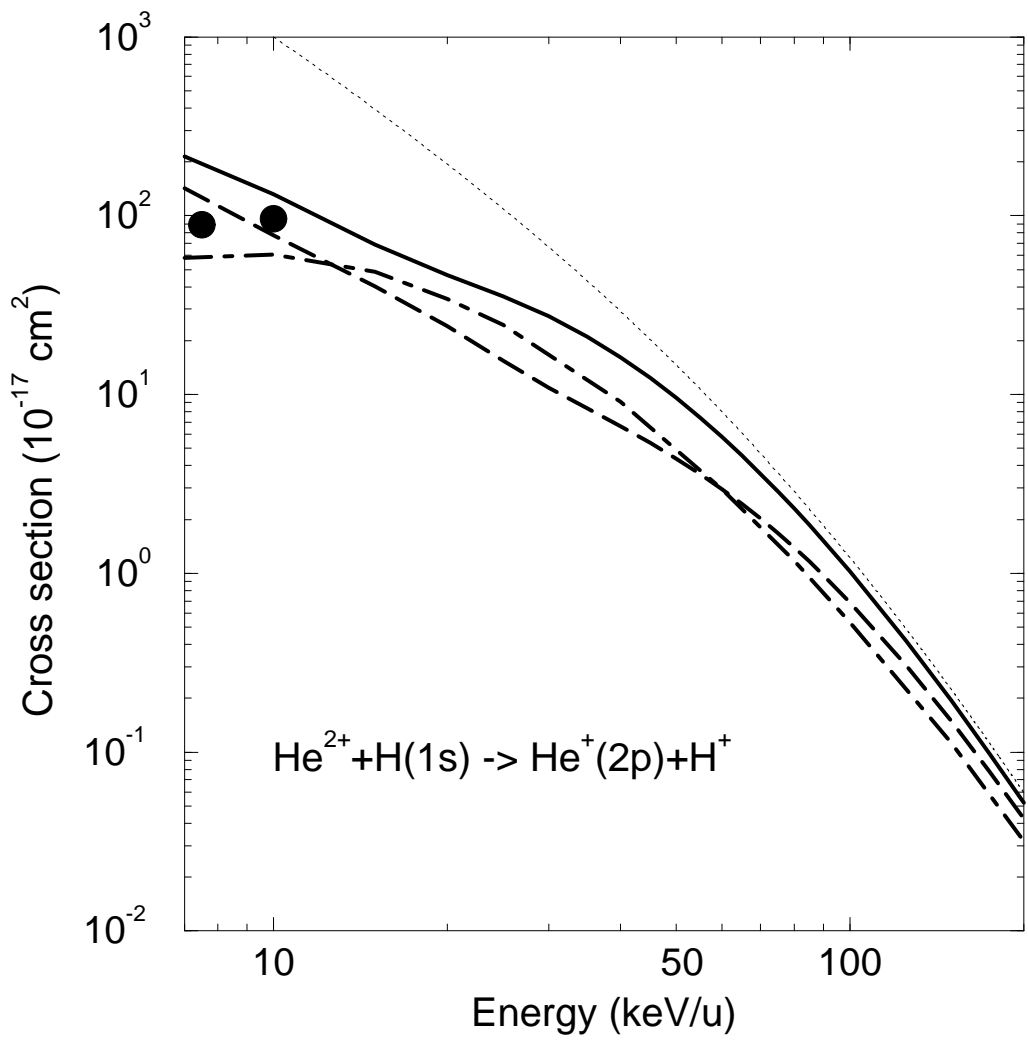
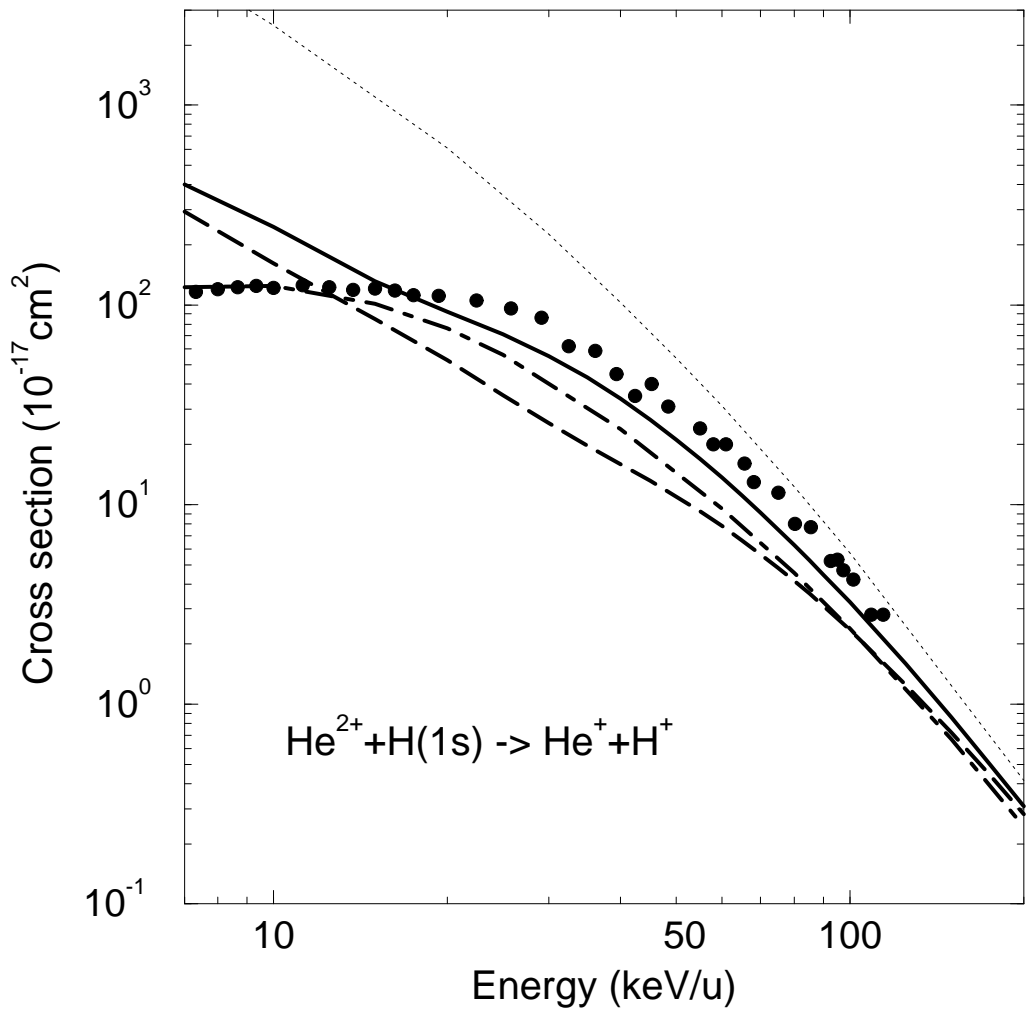
\end{document}